\documentclass[aps,showpacs,floatfix,twocolumn,byrevtex,superscriptaddress]{revtex4-1}

\usepackage[english]{babel}
\usepackage{amsmath}
\usepackage{graphicx}
\usepackage{float}
\usepackage{bm}
\usepackage{multirow}
\usepackage{dcolumn}
\usepackage{hyperref}
\usepackage[normalem]{ulem}
\usepackage[dvipsnames,usenames]{color}

\newcommand{\icm}{\ensuremath{~\textrm{cm}^{-1}}}% % cm-1
\newcommand{\BKFAx}{Ba$_{1-x}$K$_{x}$Fe$_{2}$As$_{2}$}
\newcommand{\BFAPx}{BaFe$_{2}$(As$_{1-x}$P$_{x}$)$_{2}$}
\newcommand{\BFCAx}{Ba(Fe$_{1-x}$Co$_{x}$)$_{2}$As$_{2}$}

\begin{document}

\title{Infrared probe of the gap evolution across the phase diagram of Ba$_{1-x}$K$_{x}$Fe$_{2}$As$_{2}$}
\author{B. Xu}
\affiliation{LPEM, ESPCI Paris, PSL Research University, CNRS, 10 rue Vauquelin, F-75231 Paris Cedex 5, France}
\affiliation{Beijing National Laboratory for Condensed Matter Physics, Institute of Physics, Chinese Academy of Sciences, Beijing 100190, China}

\author{Y. M. Dai}
\email[]{ymdai@lanl.gov}
\affiliation{LPEM, ESPCI Paris, PSL Research University, CNRS, 10 rue Vauquelin, F-75231 Paris Cedex 5, France}
\affiliation{Beijing National Laboratory for Condensed Matter Physics, Institute of Physics, Chinese Academy of Sciences, Beijing 100190, China}
\affiliation{Sorbonne Universit\'es, Univ Paris 06, CNRS, LPEM, F-75005 Paris Cedex 5, France}

\author{H. Xiao}
\affiliation{Center for High Pressure Science and Technology Advanced Research, Beijing 100094, China}

\author{B. Shen}
\author{H. H. Wen}
\affiliation{National Laboratory of Solid State Microstructures and Department of Physics, Nanjing University, Nanjing 210093, China}

\author{X. G. Qiu}
\affiliation{Beijing National Laboratory for Condensed Matter Physics, Institute of Physics, Chinese Academy of Sciences, Beijing 100190, China}

\author{R. P. S. M. Lobo}
\email[]{lobo@espci.fr}
\affiliation{LPEM, ESPCI Paris, PSL Research University, CNRS, 10 rue Vauquelin, F-75231 Paris Cedex 5, France}
\affiliation{Sorbonne Universit\'es, Univ Paris 06, CNRS, LPEM, F-75005 Paris Cedex 5, France}

\date{\today}

%%%%%%%%%%%%%%%%%%%%%%%%%%%%%%%%%%%%
%
% Abstract
%

\begin{abstract}
We measured the optical conductivity of superconducting single crystals of Ba$_{1-x}$K$_{x}$Fe$_{2}$As$_{2}$ with $x$ ranging from 0.40 (optimal doping, $T_c = 39$~K) down to 0.20 (underdoped, $T_c = 16$~K), where a magnetic order coexists with superconductivity. In the normal state, the low-frequency optical conductivity can be described by an incoherent broad Drude component and a coherent narrow Drude component: the broad one is doping-independent, while the narrow one shows strong scattering in the heavily underdoped compound. In the superconducting state, the formation of the condensate leads to a low-frequency suppression of the optical conductivity spectral weight. In the heavily underdoped region, the superfluid density is significantly suppressed, and the weight of unpaired carriers rapidly increases. We attribute these results to changes in the superconducting gap across the phase diagram, which could show a nodal-to-nodeless transition due to the strong interplay between magnetism and superconductivity in underdoped Ba$_{1-x}$K$_{x}$Fe$_{2}$As$_{2}$.
\end{abstract}

%  72.15.-v  Electronic conduction in metals and alloys
%  74.70.-b  SC: Superconducting materials other than cuprates
%  78.20.-e  Optical properties of bulk materials and thin films
%  78.30.-j  Infrared and Raman spectra


\maketitle

%%%%%%%%%%%%%%%%%%%%%%%%%%%%%%%%%%%%%%%%%%%%%%%%%%%%%%%%%%%%%%%%%%%%%%%%%%%%%%%
%
% Introduction
%
A fundamental question in the physics of superconductors is the nature of the electron pairing mechanism~\cite{Wang2011}. In iron-based superconductors, there are two strong candidates. One is the spin-fluctuation-mediated $s_{\pm}$ pairing state~\cite{Mazin2008,Kuroki2008}. The other is the orbital-fluctuation-mediated $s_{++}$ pairing state~\cite{Kontani2010,Kontani2011,Johnston2014}. It is not easy to assess the relative importance of orbital fluctuations and spin fluctuations since their measurable quantities often track each other closely, making the origin of pairing state remain a controversial issue~\cite{Fernandes2014}. Moreover, this issue is further complicated by the rich variety of gap structure in iron-based superconductors. The superconducting gap of systems derived from BaFe$_2$As$_2$ (Ba122) has been extensively studied~\cite{Paglione2010}. In particular, the superconducting gap in optimally (hole) doped \BKFAx\ (BKFA) was found to be quasi isotropic~\cite{Ding2008,Nakayama2009,Ren2008,Shan2011,Popovich2010,Hashimoto2009,Li2008}. However, the superconducting gap structure is highly diverse in the Ba122 families. In electron-doped \BFCAx, the gap is isotropic at the optimal doping~\cite{Luan2011}, but it develops nodes in both under- and over-doped regimes~\cite{Gordon2009,Tanatar2010,Reid2010}. In isovalent-doped \BFAPx\ the gap has nodes throughout the whole phase diagram~\cite{Hashimoto2010a,Nakai2010,Yamashita2011,Zhang2012a}. Even in BKFA, the gap value varies with $k$ for some dopings~\cite{Luo2009,Martin2009}. For example, the heavily hole-doped $x \approx 1$ compound exhibits nodal superconductivity~\cite{Hideto2009,Dong2010,Hashimoto2010b}.

% Figure 1
%
\begin{figure}[b]
\includegraphics[width=0.95\columnwidth]{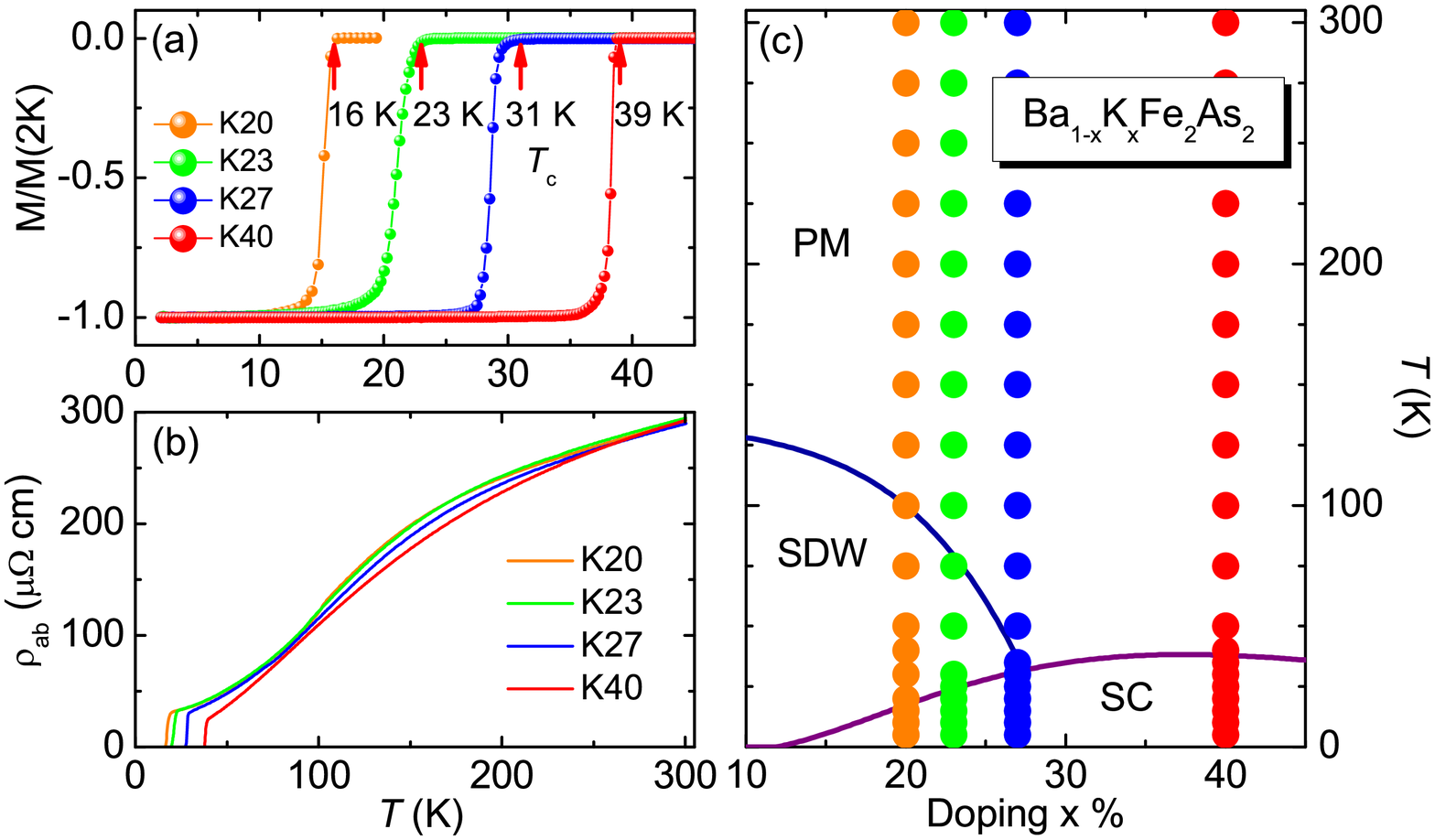}
\caption{(color online) Temperature dependence of (a) the magnetic susceptibilities of \BKFAx\ ($x$ = 0.20, 0.23, 0.27, and 0.40) in a 10 Oe magnetic field and (b) their respective in-plane resistivity. The red arrow indicates the onset of $T_c$ for each sample. (c) Phase diagram of \BKFAx\ (from Ref.~\onlinecite{Blomberg2013}). The solid circles denote the measured temperatures for each sample.}
\label{Fig1}
\end{figure}
%
% Figure 2
%
\begin{figure*}[tb]
\includegraphics[width=2\columnwidth]{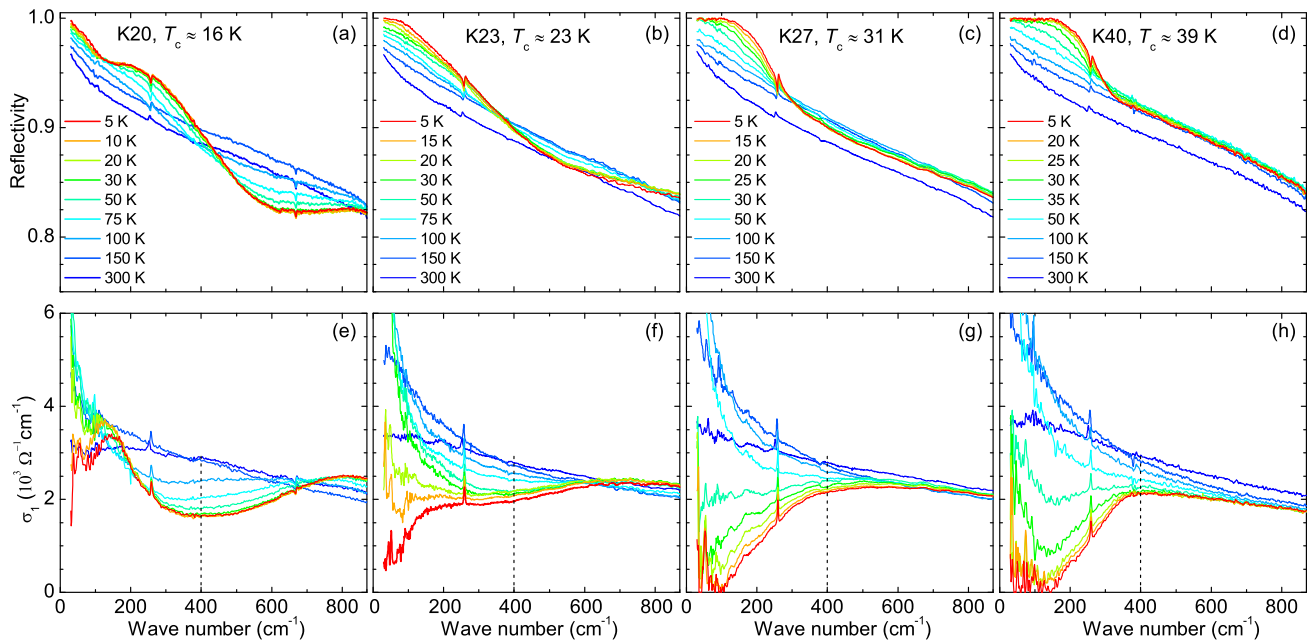}
\caption{ (color online) (a--d) Far-infrared reflectivity at several temperatures above and below $T_c$ in \BKFAx. (e--h) Temperature dependence of the corresponding optical conductivity. The vertical dashed lines indicate the cut-off frequency $\omega_c = 400\icm$ for an estimation of the weight of free carriers in the normal state.}
\label{Fig2}
\end{figure*}
The change of the gap structure in underdoped materials was ascribed to the coexistence of magnetism and superconductivity~\cite{Maiti2012}. Theoretically, the long-range magnetic order gives the coexistence states significant angular dependence through an anisotropic reconstruction of the Fermi surface, leading to a strong anisotropy in the gap function~\cite{Maiti2012}. As a result, the superconducting gap develops minima and may give rise to nodes~\cite{Maiti2012}. Furthermore, whether this coexistence is microscopic or not, has strong implications on the mechanism of superconductivity. An $s_{\pm}$ pairing symmetry is possible for such coexistence, whereas the $s_{++}$ pairing state cannot coexist with a spin-density-wave (SDW) magnetic order~\cite{Fernandes2010PRBC}.

The coexistence and competition between magnetism and superconductivity is a central point in the physics of BKFA~\cite{Yi2014,Kim2014,Cho2014,Boehmer2015,Reid2016}. Angle-resolved photoemission spectroscopy shows the coexistence of SDW and superconducting gaps as well as a dynamic competition between them~\cite{Yi2014}. Using thermal-expansion and specific-heat measurements, \citeauthor{Boehmer2015}~\cite{Boehmer2015} found an additional $C_4$-symmetric SDW phase at the coexistence region, which competes with superconductivity. \citeauthor{Kim2014}~\cite{Kim2014} observed a rapid rise of the magnetic penetration depth in underdoped compounds as a result of an anisotropic superconducting gap. Thermal conductivity measurements revealed that the reconstruction of the Fermi surface caused by the magnetic order produces a $k$-dependent gap~\cite{Reid2016}. These observations show a rich set of phenomena in the coexistence state, which have consequences on the optical response. However, most optical studies have been centered around optimally-doped BKFA~\cite{Dai2013EPL,Dai2013PRL,Li2008,Charnukha2011}. There is few systematic optical data in the underdoped region~\cite{Dai2012PRB,Mallett2016}, in particular, data related to the superconducting state.

We measured the optical conductivity of BKFA single crystals from the optimally doped material to heavily underdoped compounds. We observed isotropic superconducting gaps around optimal doping. In the heavily underdoped compounds, we found a residual low-frequency optical conductivity that takes spectral weight from the condensate, hence suppressing the superfluid density. These results can be understood in the framework of an evolution of the superconducting gap across the phase diagram, in particular with respect to the presence of nodes, caused by magnetic order in the underdoped region.

%%%%%%%%%%%%%%%%%%%%%%%%%%%%%%%%%%%%%%%%%%%%%%%%%%%%%%%%%%%%%%%%%%%%%%%%%%%%%%%
%
% Experiments
%
High quality \BKFAx\ single crystals were grown by a flux method \cite{Shen2011}. We obtained four different dopings with $x =$~0.20 (K20), 0.23 (K23), 0.27 (K27), and 0.40 (K40).  Figure~\ref{Fig1}(a) shows the magnetic susceptibility, measured in a 10~Oe magnetic field. $T_c$ was defined as the onset of the zero-field-cooling diamagnetic susceptibility, as indicated by the red arrows. In all samples, as shown in Fig.~\ref{Fig1}(b), the in-plane resistivity $\rho_{ab}(T)$ above the superconducting transition has a metallic temperature evolution.

We measured the \emph{ab}-plane reflectivity $R(\omega)$ at near-normal incidence on Bruker IFS113v and IFS66v spectrometers between 30 and $15\,000\icm$ at the temperatures indicated in Fig.~\ref{Fig1}(c). Data were collected in an Helitran ARS crysostat on a freshly cleaved surface for each sample. An \emph{in situ} gold overfilling technique~\cite{Homes1993} was used to obtain the absolute reflectivity. We extended $R(\omega)$ to $40\,000\icm$ at room temperature with an AvaSpec-$2048 \times 14$ optical fiber spectrometer. The optical conductivity $\sigma_1(\omega)$ was obtained by Kramers-Kronig analysis with either a Hagen-Rubens ($R = 1 - A\sqrt{\omega}$) or a superconducting ($R = 1 - A\omega^4$) low frequencies extrapolation. Above $15\,000\icm$ we utilized the room temperature data for all temperatures, followed by a constant reflectivity up to 100\,000 \icm\ (12.5~eV ), and a free-electron ($\omega^{-4}$) high-frequency termination.

%%%%%%%%%%%%%%%%%%%%%%%%%%%%%%%%%%%%%%%%%%%%%%%%%%%%%%%%%%%%%%%%%%%%%%%%%%%%%%%
%
% Data analysis
%
Figures~\ref{Fig2}(a--d) show the far-infrared $R(\omega)$ at several temperatures above and below $T_c$ for all samples. In the paramagnetic state, $R(\omega)$ has a typical metallic response with a high reflectivity that increases upon cooling. Upon entering the SDW state, the reflectivity spectra of K20 and K23 are renormalized around 400\icm\ due to the opening of the SDW gap~\cite{Hu2008,Dai2012PRB}. In the superconducting state, all samples show an upturn in $R(\omega)$ at low frequencies, which is a consequence of the formation of the superconducting condensate. In K40 and K27, $R(\omega)$ at 5~K has a sharp edge and rises to a flat 100\% value expected for an $s$-wave superconductor, suggesting an isotropic superconducting gap. In K23 and K20, $R(\omega)$ at 5~K does not reach a flat response, indicating the presence of unpaired quasiparticles.

Figures~\ref{Fig2}(e--h) show the temperature dependence of $\sigma_1(\omega)$ in the far-infrared region for all samples. In the paramagnetic state, the low-frequency $\sigma_1(\omega)$ has a Drude-like peak. In the underdoped samples, the Fermi surface gets partially gaped below $T_{SDW}$~\cite{Hu2008,Dai2012PRB}, showing a characteristic transfer of spectral weight from frequencies below 650\icm, to values above this energy. The formation of the superconducting condensate below $T_c$ induces a suppression of $\sigma_1(\omega)$ at low frequencies in all samples. This missing spectral weight is transferred to the $\delta(\omega)$ function representing the infinite dc conductivity.

% Figure 3
%
\begin{figure}[tb]
\includegraphics[width=\columnwidth]{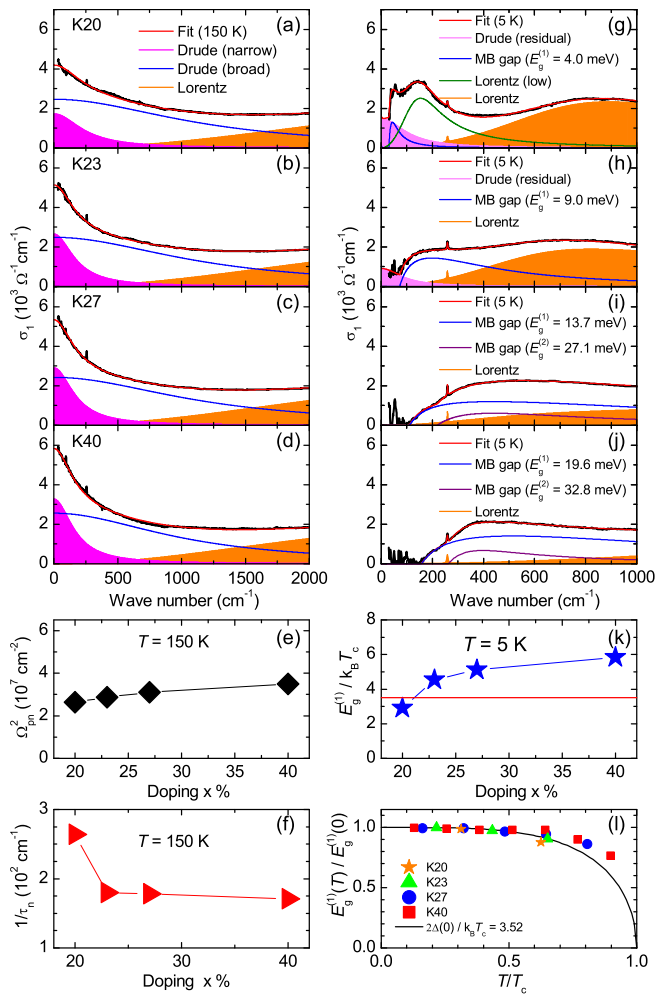}
\caption{ (color online) Optical conductivity at $T$ = 150~K for (a) K20, (b) K23, (c) K27 and (d) K40. Each spectrum is decomposed into individual contributions of a Drude-Lorentz approach. Doping dependence of (e) the Drude weight $\Omega_{pn}^2$ and (f) scattering rate $1/\tau_n$ for the narrow Drude term. Optical conductivity at $T$ = 5~K for (g) K20, (h) K23, (i) K27 and (j) K40. To decompose the total response into individual contributions in the superconducting state, some Drude peaks were replaced by Mattis-Bardeen terms. $E^{(1)}_g$ and $E^{(2)}_g$ are the obtained values of the superconducting gaps. (k) Doping dependence of the ratio $E^{(1)}_g/k_BT_c$. The red line indicate the value of BCS weak coupling limit. (l) Normalized superconducting gap $E_g^{(1)}(T)/E_g^{(1)}(0)$ versus $T/T_c$ for all samples. The solid black line is a BCS calculation.}
\label{Fig3}
\end{figure}
The normal state $\sigma_1(\omega)$ of pnictides can be conveniently parameterized by a Drude-Lorentz model~\cite{Wu2010,Tu2010,Nakajima2010,Dai2013PRL,Dai2015PRX}:
\begin{equation}
\label{DrudeLorentz}
\sigma_{1}(\omega)=\frac{2\pi}{Z_{0}}\! \left[
  \sum_k\frac{\Omega^{2}_{p,k}}{\tau_k(\omega^{2}+\tau_k^{-2})}
  \!+\!
  \sum_j\frac{\gamma_j\omega^{2}S_j^{2}}{(\omega_j^{2}-\omega^{2})^{2}+\gamma_j^{2}\omega^{2}}
  \right]
\end{equation}
where $Z_{0}$ is the vacuum impedance. The first term describes a sum of free-carrier Drude responses, each characterized by a plasma frequency $\Omega_{p}^2 = ne^2/\varepsilon_0 m^\ast$, where $n$ is the carrier concentration, $m^\ast$ the effective mass, and $1/\tau$ the scattering rate. The second term is a sum of Lorentz oscillators, each having a resonance frequency $\Omega_j$, a line width $\gamma_j$ and a weight $S_j$.

Figures~\ref{Fig3}(a--d) show the measured $\sigma_{1}(\omega)$ (black curves) and corresponding fits to the data at 150~K using Eq.~\ref{DrudeLorentz} for all four samples. For each compound, the fitting curve is decomposed into a narrow Drude term (magenta line) and a broad Drude term (blue line), associated to the multi-band feature of iron-based superconductors, as well as a Lorentz term (orange line) describing the interband transitions in the high energy region. All $\sigma_1(\omega)$ curves in the paramagnetic state can be fitted with equal quality by using this approach, and the corresponding fitting parameters are summarized in Table~\ref{Table1}. As shown by its parameters, the broad Drude term (D2) produces an incoherent background contribution without much temperature and doping dependence. On the contrary, the narrow Drude term is strongly temperature and doping dependent. For example, in K40, it features a $T$-linear scattering rate that originates from the presence of a quantum critical point in the superconducting dome~\cite{Dai2013PRL}. Here, we focus on the doping dependence of this narrow Drude term. As shown by the black diamonds in Fig.~\ref{Fig3}(e), the Drude weight $\Omega^2_{pn}$ increases slightly with K doping, which is also revealed by the magenta filled area in Fig.~\ref{Fig3}(a--d). Meanwhile, the hole doping drives the system towards the $3d^5$ state where Hund's rule coupling is the strongest, giving an enhancement of the carrier effective mass $m^{\ast}$~\cite{Medici2016,Ye2014}. However, one should note that the hole density steadily increases with K doping~\cite{Shen2011,Malaeb2012,Ye2014}. As a result, because of a balancing between $n$ and $m^{\ast}$, $\Omega^2_{pn}$ shows only a slight doping dependence. It is also worth noticing the doping dependence of the scattering rate $1/\tau_{n}$, as shown in Fig.~\ref{Fig3}(f), which shows much stronger scattering around the doping K20. In this region, as we discussed later, a stronger suppression of superfluid density was also observed, suggesting a close connection between the normal state and the superconducing state.

%%%%%%%%%%%%%%%%%%%%%%%%%%%%%%%%%
%Table 1
\begin{table*}[t]
\caption{\label{Table1}%
The results of the nonlinear least-squares fit of the Drude-Lorentz model and the Mattis-Bardeen model to the optical conductivity of \BKFAx\ at several temperatures from 300 to 5~K. The terms D1 and D2 denote the two Drude contributions, while L1 is the Lorentz oscillator for the high-energy interband transitions; the contribution of the SDW gap that appears below $T_{SDW}$ is denoted by the oscillator L02; the oscillator L01 repesents the new low-energy excitations in K20; D$_{res}$ is the residual Drude contribution from the unpaired quasiparticles in the superconducting state; $E^{(2)}_g$ and $E^{(1)}_g$ are the values of the large and small supercondcuting gaps obtained by the Mattis-Bardeen model. The estimated errors for all parameters are 5\% or less. All units are in \icm.}
\begin{ruledtabular}
\begin{tabular}{cccccccccccccccc}
\multicolumn{16}{c}{(a) K20 ($T_c \simeq$ 16 K)}\\
\hline
& \multicolumn{3}{c}{D1/D$_{res}$} & \multicolumn{3}{c}{D2/MB gap} & \multicolumn{3}{c}{L01} & \multicolumn{3}{c}{L02} & \multicolumn{3}{c}{L1} \\
\cline{2-4}\cline{5-7}\cline{8-10}\cline{11-13}\cline{14-16}
$T$ (K) & $\Omega_{p,D1}$  & $1/\tau_1$ &   &$\Omega_{p,D2}$  & $1/\tau_2$ & $E^{(1)}_g$ &$\omega_{01}$  & $\gamma_{01}$ & $S_{01}$ &$\omega_{02}$  & $\gamma_{02}$ & $S_{02}$ &$\omega_{1}$  & $\gamma_1$ & $S_1$ \\
\hline
300  & 5144 & 686 &  & 13190 & 1151 &  &  &  &  &  &  &  & 5615 & 14443 & 47484\\
250  & 5096 & 442 &  & 13165 & 1175 &  &  &  &  &  &  &  & 5643 & 13518 & 45624\\
200	 & 5144	& 329 &  & 13195 & 1179 &  &  &  &  &  &  &  & 5750 & 13001 & 45182\\
150	 & 5126	& 264 &  & 13120 & 1164 &  &  &  &  &  &  &  & 5826 & 12657 & 45068\\
100	 & 4669	& 103 &  & 8813  & 712 &  &  &  &  & 742 & 1222 & 9436 & 5915 & 12605 & 45125\\
50	 & 4205	& 22  &  & 3767 & 133 &  & 133 & 254 & 5684 & 819 & 1052 & 11152 & 5901	& 12374 & 45071\\
5    & 3114 & 109 &  & 3122 & 28 & 32 & 154 & 203 & 5530 & 853 & 1030 & 11340 & 5919 & 12396 & 45276\\
\hline \\
\multicolumn{16}{c}{(b) K23 ($T_c \simeq$ 23 K)}\\
\hline
& \multicolumn{3}{c}{D1/D$_{res}$} & \multicolumn{3}{c}{D2/MB gap} & \multicolumn{3}{c}{L01} & \multicolumn{3}{c}{L02} & \multicolumn{3}{c}{L1} \\
\cline{2-4}\cline{5-7}\cline{8-10}\cline{11-13}\cline{14-16}
$T$ (K) & $\Omega_{p,D1}$  & $1/\tau_1$ &   &$\Omega_{p,D2}$  & $1/\tau_2$ & $E^{(1)}_g$ &$\omega_{01}$  & $\gamma_{01}$ & $S_{01}$ &$\omega_{02}$  & $\gamma_{02}$ & $S_{02}$ &$\omega_{1}$  & $\gamma_1$ & $S_1$ \\
\hline
300	 & 5329	& 457 & & 13113	& 1222 &  &  &  &  &  &  &  & 5511	& 14914	& 47680\\
250	 & 5300	& 312 & & 13113	& 1205 &  &  &  &  &  &  &  & 5620	& 14919	& 47710\\
200	 & 5324	& 243 & & 13143	& 1166 &  &  &  &  &  &  &  & 5690	& 14420	& 47178\\
150	 & 5351	& 180 & & 13149	& 1157 &  &  &  &  &  &  &  & 5813	& 14080	& 46956\\
100	 & 5338	& 98  & & 13141	& 1196 &  &  &  &  &  &  &  & 5838	& 13856	& 46775\\
50	 & 5147	& 26  & & 7161 & 266   &  &  &  &  & 718 & 1128 & 10866 & 5777	& 13636 & 46653\\
5    & 2230 & 91 &  & 7090 & 392 & 72&  &  &   & 757 & 1056 & 10134 & 5766 & 13725 & 49801\\
\hline \\
\multicolumn{16}{c}{(c) K27 ($T_c \simeq$ 31 K)}\\
\hline
& \multicolumn{3}{c}{D1/MB gap} & \multicolumn{3}{c}{D2/MB gap} & \multicolumn{3}{c}{L01} & \multicolumn{3}{c}{L02} & \multicolumn{3}{c}{L1} \\
\cline{2-4}\cline{5-7}\cline{8-10}\cline{11-13}\cline{14-16}
$T$ (K) & $\Omega_{p,D1}$  & $1/\tau_1$ & $E^{(2)}_g$  &$\Omega_{p,D2}$  & $1/\tau_2$ & $E^{(1)}_g$ &$\omega_{01}$  & $\gamma_{01}$ & $S_{01}$ &$\omega_{02}$  & $\gamma_{02}$ & $S_{02}$ &$\omega_{1}$  & $\gamma_1$ & $S_1$ \\
\hline
300	& 5493	& 449 &  & 13137 & 1263	&  &  &  &  &  &  &  & 5545 & 15223 & 48381\\
250	& 5504	& 336 &  & 13056 & 1231	&  &  &  &  &  &  &  & 5694 & 15282 & 48669\\
200	& 5504	& 260 &  & 13063 & 1177	&  &  &  &  &  &  &  & 5728 & 14612 & 47868\\
150	& 5578	& 178 &  & 13111 & 1181	&  &  &  &  &  &  &  & 5862 & 14369 & 47744\\
100	& 5512	& 87  &  & 13136 & 1139	&  &  &  &  &  &  &  & 5904 & 13965 & 47404\\
50	& 5510	& 29  &  & 12403 & 1071	&  &  &  &  & 658 & 1305 & 5203 & 5861 & 13028 & 45993\\
5   & 6257 & 559 & 218 & 10718 & 1303 & 110 &  &  &  & 700 & 1485 & 6611 & 5942 & 12830 & 45641\\
\hline \\
\multicolumn{16}{c}{(d) K40 ($T_c \simeq$ 39 K)}\\
\hline
& \multicolumn{3}{c}{D1/MB gap} & \multicolumn{3}{c}{D2/MB gap} & \multicolumn{3}{c}{L01} & \multicolumn{3}{c}{L02} & \multicolumn{3}{c}{L1} \\
\cline{2-4}\cline{5-7}\cline{8-10}\cline{11-13}\cline{14-16}
$T$ (K) & $\Omega_{p,D1}$  & $1/\tau_1$ & $E^{(2)}_g$  &$\Omega_{p,D2}$  & $1/\tau_2$ & $E^{(1)}_g$ &$\omega_{01}$  & $\gamma_{01}$ & $S_{01}$ &$\omega_{02}$  & $\gamma_{02}$ & $S_{02}$ &$\omega_{1}$  & $\gamma_1$ & $S_1$ \\
\hline
300  & 5842 & 405 &  & 12908 & 1159 &  &  &  &  &  &  &  & 5901 & 17741 & 50703\\
250  & 5865 & 328 &  & 12908 & 1145 &  &  &  &  &  &  &  & 5949 & 17407 & 50344\\
200	 & 5936	& 250 &  & 12920 & 1134 &  &  &  &  &  &  &  & 5999 & 16556 & 49467\\
150	 & 5918	& 171 &  & 12961 & 1117 &  &  &  &  &  &  &  & 6063 & 15842 & 48656\\
100	 & 5900	& 95  &  & 12919 & 1122 &  &  &  &  &  &  &  & 6114 & 15312 & 48046\\
50	 & 5910	& 28  &  & 12947 & 1117 &  &  &  &  &  &  &  & 6132	& 14920 & 47699\\
5    & 6652 & 281 & 263 & 11885 & 1262 & 157&  &  &  &  &  &  & 6205 & 15927 & 49331\\
\end{tabular}
\end{ruledtabular}
\end{table*}
In the superconducting state, we parametrized the data by replacing the Drude term with a Mattis-Bardeen (MB) term~\cite{Mattis1958,Zimmermann1991}. For K40 and K27, as shown by the fitting results of data at 5~K in Fig.~\ref{Fig3}(j) and Fig.~\ref{Fig3}(i), the best fit requires two MB terms to describe the contributions of superconducting gaps. These two MB terms have absorption edges at $E^{(1)}_{g}$ = 19.6~meV and $E^{(2)}_{g}$ = 32.8~meV for K40, and $E^{(1)}_{g}$ = 13.7~meV and $E^{(2)}_{g}$ = 27.1~meV for K27. For K23 [Fig.~\ref{Fig3}(h)] and K20 [Fig.~\ref{Fig3}(g)], a Lorentz term is required to depict the SDW gap around 800\icm, and a Drude term is kept to describe the residual $\sigma_1(\omega)$ existing at low frequencies. The absorption edge of the superconducting gap is modeled by a single MB term at $E^{(1)}_{g} =$ 9.0 and 4.0~meV for K23 and K20, respectively. The fitting results both in the SDW state and the superconducting state are also summarized in Table~\ref{Table1}. More intriguingly, in K20, an additional Lorentz term around 150\icm\ sets in, implying the emergence of low-energy transitions, as observed in Co-doped Ba122~\cite{Lobo2010,vanHeumen2010}. This low-energy feature and the residual Drude response substantially deplete the superfluid weight.

% Figure 4
%
\begin{figure}[tb]
\includegraphics[width=0.9\columnwidth]{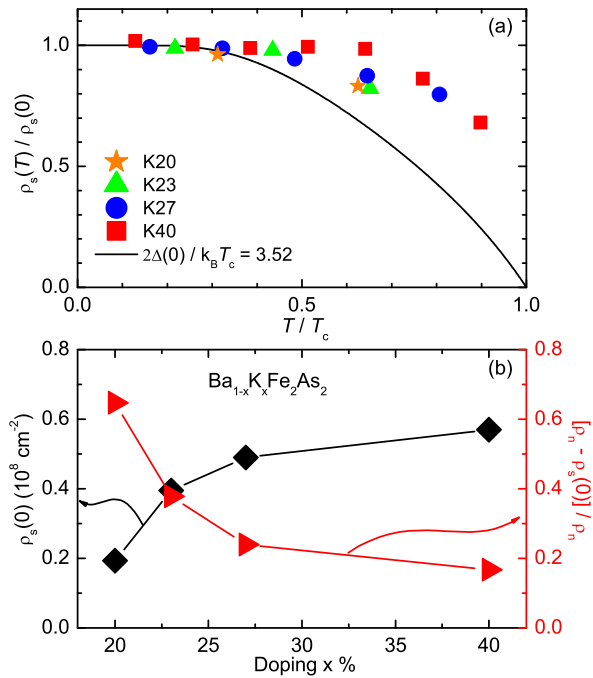}
\caption{ (color online) (a) Normalized superfluid density $\rho_{s}(T)/\rho_{s}(0)$ versus $T/T_c$. The solid lines are BCS calculations. (b) Doping dependence of $\rho_s(0)$ (black diamonds) and $[\rho_n - \rho_s(0)]/\rho_n$ (red triangles).}
\label{Fig4}
\end{figure}
Before examining the superfluid weight, we first discuss the strength of the electron-boson coupling in BKFA. Here, the utilization of the Mattis-Bardeen formalism assummes a weakly-coupled superconductor and the absorption edge occurs at $E_g = 2 \Delta$. In strongly coupled pnictides~\cite{Benfatto2009,Popovich2010,Hardy2016,Charnukha2011,Dai2013EPL}, a multiband Eliashberg analysis is more accurate~\cite{Charnukha2011} and the absorption edge happens at $E_g = E_B + 2\Delta$, where $E_B$ is the energy of the exchange boson (the so-called Holstein process)~\cite{Dai2013EPL,Akis1991}. This is indeed the case for the samples K40 and K27, in which the gap values, especially $E^{(2)}_{g}$, are higher than the reported range for these materials~\cite{Ding2008,Nakayama2009,Ren2008,Shan2011,Popovich2010}. The reasonable explanation is in these dopings the coupling is strong, and the absorption of the gap is coupled with a boson mode energy $E_{B}$~\cite{Christianson2008,Wang2012}. However, Mattis-Bardeen may remain a valid approximation for the small gap $E^{(1)}_{g}$, since it only associates the energy where $\sigma_1(\omega)$ starts to rise. In addition, in the lower dopings, the coupling becomes weaker. As shown in Fig.~\ref{Fig3}(k), the estimated $E^{(1)}_g/k_{B}T_{c}$ depends on the doping and shows a strong coupling value for the optimal doping K40, which drops to a value below the BCS weak coupling limit for the underdoped K20. In Fig.~\ref{Fig3}(l), the temperature dependence of $E^{(1)}_{g}(T)/E^{(1)}_{g}(0)$ also supports this doping evolution, in which the values for K40 and K27 are above the simple BCS calculation, suggesting a stronger coupling.

Let us now discuss the superfluid weight in BKFA, obtained from the finite frequency $\sigma_1$ spectral weight lost below $T_c$ or, equivalently, the $1/\omega$ contribution expected in the imaginary optical conductivity~\cite{Dordevic2002,Zimmers2004}. Figure~\ref{Fig4}(a) shows $\rho_{s}(T)/\rho_{s}(0)$ as a function of temperature for all samples. The solid line is the solution of the BCS penetration depth equation with $2\Delta(0)/k_BT_c = 3.52$.

The doping dependence of $\rho_{s}(0)$ is displayed in Fig.~\ref{Fig4} (b) as black diamonds. It decreases monotonically with decreasing $x$, exhibiting a substantial drop at the lowest doping. One should note that the residual Drude response in the more underdoped materials is a signature of unpaired carriers that take spectral weight away from the superfluid. To compare $\rho_{s}(0)$ and the weight of unpaired carriers, we calculated the ratio $[\rho_{n} - \rho_{s}(0)]/\rho_{n}$, where $\rho_{n}$ is an estimation of the weight for the free carriers in the normal state, defined as
\begin{equation}
\rho_{n}(\omega_c, T) \equiv \frac{Z_0}{\pi^2}\int_{0}^{\omega_c}\sigma_1(\omega, T)d\omega,
\end{equation}
where $Z_0$ is the vacuum impedance. We took $\rho_{n}$ at $T = 50$~K with a cut-off frequency $\omega_c = 400\icm$ (indicated by the vertical dashed lines in Fig.~\ref{Fig2}). The doping dependence of the ratio $[\rho_{n} - \rho_{s}(0)]/\rho_{n}$ is plotted in Fig.~\ref{Fig4}(b) as red triangles. This estimation shows that most of the carriers in the optimal doping K40 are condensed. With decreasing $x$, $[\rho_{n} - \rho_{s}(0)]/\rho_{n}$ increases gradually and rises steeply in the heavily underdoped samples, implying a significantly suppressed superfluid condensate.

In order to understand the doping evolution of the superconducting response in the underdoped BKFA, we consider the competition and coexistence between magnetism and superconductivity. The SDW magnetic order competes with superconductivity for the same electronic states~\cite{Chubukov2009PRB,Fernandes2010,Mallett2016}. As shown in Figs.~\ref{Fig3}(g--h), the SDW gap depletes the low-energy spectral weight available for the superconducting condensate. As a result, the superfluid density is suppressed as the SDW order sets in. In a different perspective, the coexistence of superconductivity with a striped SDW magnetism leads to a strong anisotropy in the gap function~\cite{Maiti2012}. Theoretically, in the SDW magnetic phase, the new magnetic Brillouin zone boundary would reconstruct the Fermi surface due to a unit cell doubling~\cite{Maiti2012,Reid2016}. \citeauthor{Maiti2012} proposed that the reconstructed Fermi surface is anisotropic and shrinks gradually as the SDW order increases, giving rise to a strong modulation on the gap structure~\cite{Maiti2012}. As a result, the gap develops an anisotropy that can lead to a nodeless-nodal transition in the coexistence region. The following sequence naturally explains our results: around optimal doping, the superconducting gaps are almost isotropic, i.e. nodeless. Decreasing $x$ to the onset of the coexistence region, the gap anisotropy increases but remains nodeless. Moving deeper into the coexistence region, the gap becomes strongly anisotropic, possibly with nodes, and residual unpaired carriers are present.

Furthermore, the strong reconstruction of the Fermi surface would induce deep minima or even an energy gap at the crossing points of the Brillouin zone~\cite{Maiti2012,Reid2016}. Thus, the low-energy peak in K20 is likely related to the excitations from such deep minima or energy gap. These excitations further share the electronic states available for the superconducting condensate, leading to the greater suppression of superfluid density in the coexistence region. At the boundary between striped SDW magnetism and superconductivity in the phase diagram of BKFA, an additional $C_4$-symmetric SDW phase has been observed in thermodynamic experiments~\cite{Boehmer2015}. This $C_4$ state reduces the anisotropy around the boundary region. As a result, for K23 and K27, this low-energy absorption peak is absent and the suppression of superfluid density is relatively small.

%%%%%%%%%%%%%%%%%%%%%%%%%%%%%%%%%%%%%%%%%%%%%%%%%%%%%%%%%%%%%%%%%%%%%%%%%%%%%%%
%
% Conclusions
%
To summarize, we performed a systematic optical study in \BKFAx\ from the optimally doped to heavily underdoped regimes. Near optimal doping, an isotropic and nodeless superconducting gap was observed. As the doping level $x$ decreases, a residual Drude response and low-energy excitations emerge, suggesting the existence of unpaired quasiparticles in the region where magnetism and superconductivity coexist. Meanwhile, as the magnetic order is enhanced, the superfluid density is strongly suppressed and the scattering of carriers increases. These observations suggest that the competition and coexistence between magnetic order and superconductivity play an important role in the doping evolution of superconducting gap structure in \BKFAx, indicating a nodeless to nodal transition with decreasing doping.

%%%%%%%%%%%%%%%%%%%%%%%%%%%%%%%%%%%%%%%%%%%%%%%%%%%%%%%%%%%%%%%%%%%%%%%%%%%%%%%
%
% Acknowledgment
%
We acknowledge discussions with A. Chubukov. Work at IOP CAS was supported by MOST (973 Projects No. 2015CB921303, and 2015CB921102), and NSFC (Grants No. 91421304, and 11374345). H. Xiao is supported by NSFC, Grant No. U1530402.
%%%%%%%%%%%%%%%%%%%%%%%%%%%%%%%%%%%%%%%%%%%%%%%%%%%%%%%%%%%%%%%%%%%%%%%%%%%%%%%
%
% The bibliography (BibTeX)
%

\end{document}